\documentclass[11pt]{article}
\usepackage{amsmath,epsf,amssymb,latexsym,enumerate,cite,shadow,array,color,bbold}
\usepackage{slashed}
\usepackage{graphicx,wasysym}

\DeclareFontFamily{U}{rsf}{} \DeclareFontShape{U}{rsf}{m}{n}{
  <5> <6> rsfs5 <7> <8> <9> rsfs7 <10-> rsfs10}{}
\DeclareMathAlphabet\Scr{U}{rsf}{m}{n} \makeatletter
\@addtoreset{equation}{section} \makeatother

\def\be{\begin{equation}}
\def\ee{\end{equation}}
\def\ba{\begin{array}}
\def\ea{\end{array}}

\newcommand{\bea}{\begin{eqnarray}}
\newcommand{\eea}{\end{eqnarray}}
\def\N{$\cal N$}

\def\cG{{\Scr G}}

\def\K{K{\"a}hler}
\def\rmd{{\rm d}}

\usepackage{ifpdf}
\ifx\pdfoutput\undefined
   \pdffalse
\else
   \pdfoutput=1
   \pdftrue
  \usepackage[pdftex]{hyperref}
\def\u0{{\underline 0}}

\def\url{{\underline {r+\ell}}}

\newcommand{\rf}[1]{(\ref{#1})}

\def\Tr{\mathop{\rm Tr}\nolimits}

\def\rmd{{\rm d}}

\newcommand{\betaplus}{\beta _+}
\newcommand{\betaminus}{\beta _-}

\begin{document}

\begin{titlepage}

\begin{flushright}
 SU/ITP-14/26
\end{flushright}

\hskip 1cm

\vskip 2cm

\begin{center}

{\LARGE \textbf{  Emergence of Spontaneously Broken Supersymmetry on an Anti-D3-Brane in KKLT dS vacua\\[0pt]
\vskip 0.8cm }}

\
 {\bf Renata Kallosh}\   {\bf and  Timm Wrase} \vskip 0.5cm
{\small\sl\noindent
 Department of Physics, Stanford University, Stanford, CA
94305 USA}
\end{center}
\vskip 1 cm

\begin{abstract}

\

\

The KKLT construction of de Sitter vacua includes an uplifting term coming from an anti-D3-brane. Here we show how this term can arise via spontaneous breaking of supersymmetry, based on the emergence of a nilpotent chiral supermultiplet on the world-volume of the anti-D3-brane.  We establish and use the fact that both the DBI as well as the WZ term, with account of orientifolding, acquire a form of the Volkov-Akulov action. For an O3 orientifold involution of $\mathbb{R}^{9,1}$ we demonstrate the cancellation between the fermionic parts of the DBI and WZ term for the D3-brane action. For the anti-D3-brane we show that the DBI action and the WZ action combine and lead to the emergence of the goldstino multiplet which is responsible for spontaneous breaking of supersymmetry. This provides a string theoretic explanation for the supersymmetric uplifting term in the KKLT effective supergravity model supplemented by a nilpotent chiral multiplet.
\end{abstract}

\vspace{24pt}
\end{titlepage}


\newpage

\section{Introduction}

The manifestly supersymmetric effective $d=4$ supergravity action describing the KKLT model of the AdS stabilization of the volume modulus in type IIB string theory results from the following \K\, potential and superpotential, \cite{Kachru:2003aw,Kachru:2003sx}:
\be
W=W_0 + Ae^{-a\rho}\, , \qquad K= - 3 \ln(\rho + \overline{\rho}) \ .
\ee
The supersymmetric AdS vacua  in KKLT models are defined by the equation $D_\rho W=0$.
The uplifting term was added in the next step in the KKLT construction in the form  \cite{Kachru:2003aw,Kachru:2003sx}
\be
\delta V= {D\over (\rho+\bar \rho)^3}\ .
\label{Vuplift}
\ee
In the string theory model \cite{Kachru:2002gs} it has been argued that the presence of the anti-D3-branes breaks supersymmetry spontaneously since the anti-D3-branes can decay to a supersymmetric state by annihilating with fluxes. However, it was not clear how to write down an effective \N=1 supergravity action: in \cite{Kachru:2003aw,Kachru:2003sx} eq. \rf{Vuplift} was used, which corresponds to a pure bosonic term breaking supersymmetry explicitly. 

It was explained in \cite{Kachru:2003sx} that 
for a  D3-brane slowly moving in the background with no anti-branes the net force vanishes due to gravitational and five-form cancellations: the relevant parts of the DBI and the WZ terms cancel.  For the anti-D3-brane the force exerted by gravity and the five-form field are of the same sign and add, so we have a factor of 2 for the anti-D3-brane versus 0 for the D3-brane, leading to \rf{Vuplift}. This argument was developed in \cite{Kachru:2003sx} in the absence of the fermions on the brane. In this paper we will find that when the fermions on the brane are taken into account and supersymmetry is broken spontaneously, the same effect, doubling versus cancellation,  of the full Volkov-Akulov goldstino action \cite{Volkov:1973ix} takes place. This will provide us with a supersymmetric uplifting of the   supergravity KKLT models which has an origin in the supersymmetric D-brane physics.

Recently,  a systematic construction of metastable de Sitter vacua in a broad class of string theory motivated supergravity models was performed in  \cite{Kallosh:2014oja}. It   has confirmed the standard expectation that supersymmetry is an indispensable tool, which helps to find many  dS vacua and simultaneously ensures their local stability.

More recently it was pointed out in \cite{Ferrara:2014kva} that in supergravity one could have started with the following supersymmetric model, depending on 2 supermultiplets, $\rho$ and $S$, where $S$ represents a   Volkov-Akulov goldstino multiplet \cite{Volkov:1973ix}
\be
W=W_0 + Ae^{-a\rho}- M^2 S\, , \qquad K= - 3 \ln(\rho + \overline{\rho})+ S\bar S\,  \qquad {\rm at} \qquad S^2=0 \ .
\label{ModelN}
\ee
Here $S$ is the nilpotent\footnote{The chiral multiplet $S(x, \theta)$ was defined off-shell in \cite{Komargodski:2009rz}. In earlier versions in \cite{rocek} in addition to the $S^2=0$ constraint, also a specific on-shell constraint was used. For cosmological applications we use the off-shell construction in \cite{Komargodski:2009rz}. } chiral supermultiplet \cite{rocek,Komargodski:2009rz}  which provides a manifestly supersymmetric version of the Volkov-Akulov goldstino. After computing the potential,  we have to set the scalar part of the superfield $S$ to zero.
We find
\be
V= V_{KKLT} (\rho, \bar \rho) + {M^4 \over (\rho+\bar \rho)^3}   \ ,
\label{potentialNew}
\ee
where $V_{KKLT} (\rho, \bar \rho)$ is the KKLT potential without the uplifting term, at $M=0$.
This shows that \rf{ModelN} corresponds to a manifestly supersymmetric supergravity version of the uplifting term arising from an anti-D3-brane (extending the  bosonic expression for the uplifting term from the anti-D3-brane used in \cite{Kachru:2003aw,Kachru:2003sx}).

For simplicity we  consider here the case without warping. This will allow us to study  the supersymmetry upon gauge-fixing of $\kappa$-symmetry on the world-volume of the brane in a flat type IIB supergravity background, which is a  relatively simple case. Generalization to a generic type IIB background will be a next step.

From the point of view of $d=4$ supergravity, the supersymmetrization of the uplifting due to a chiral nilpotent multiplet is obvious. It is less obvious how all this is related to D-brane physics and to the fact that adding a D3-brane to the system considered in \cite{Kachru:2003aw,Kachru:2003sx} will not lead to an uplift, whereas adding an anti-D3-brane, will result in the emergence of a VA multiplet and supersymmetric uplifting.

Below we will present a refined relation between our $d=4$ supergravity and D$p$-brane physics with global supersymmetry and local $\kappa$-symmetry \cite{Cederwall:1996pv, Aganagic:1996nn, Bergshoeff:1997kr, Kallosh:1997aw, Kamimura:1997ju,Bergshoeff:2013pia, Simon:2011rw}. In \cite{Ferrara:2014kva} we referred to the well known argument \cite{Kallosh:1997aw,Bergshoeff:2013pia} that a D$p$-brane action, when gauge-fixed in a certain gauge, always leads to a DBI term which has Volkov-Akulov fermions on its world volume. Not surprisingly, the non-linear VA fermions are superpartners of the Born-Infeld non-linear vectors. In the same gauge the WZ terms vanishes, as was first established in \cite{Aganagic:1996nn}. It appears therefore that the emergence of the VA fermions takes place independently of the charge of the brane: for the D$p$-brane and for the anti-D$p$-brane we are always getting the VA fermions.

However, this is not expected to be true in the context of the KKLT model, where by construction, only the anti-D3-brane can be responsible for the uplifting, a D3-brane will not do the job. There must be a reason why the emergence of the VA fermion on the word-volume of the brane is different for a D3-brane and an anti-D3-brane. And indeed, as we show in this paper, for a large class of models (that include the KKLT scenario) such a reason exists: In order to preserve $\mathcal{N}=1$ supersymmetry in $d=4$ starting from type II $\mathcal{N}=2$ supersymmetry in $d=10$ one has to compactify the theory on a Calabi-Yau manifold, and in addition perform an orientifold projection. However, the standard $\kappa$-symmetry gauge \cite{Aganagic:1996nn} in which the WZ term for any brane vanishes, is incompatible with an orientifold projection. If, instead, one uses a $\kappa$-symmetry gauge fixing that is consistent with the orientifold projection, then the WZ action does not vanish and the emergence/vanishing of the VA fermions on the world-volume indeed depends on the charge of the brane.

The remarkable discovery of the fact that the WZ term of the D9-brane with the type I orientifold truncation becomes a Volkov-Akulov goldstino action was made in \cite{Bergshoeff:1999bx,Riccioni:2003ga}. Therefore, depending on the choice of the charge of the brane, for a given choice of the sign in the orientifolding condition the total action either vanishes or becomes the sum of the two VA actions. This gives a hint on a possible reason for an analogous dependence on the charge of the brane for a D3-/anti-D3-brane in the presence of an O3 orientifold projection.

In this paper we perform a generic analysis of the D$p$-brane in a flat background and show that the WZ term upon orientifolding becomes exactly the VA action. This gives an analytic explanation of the computational result in \cite{Bergshoeff:1999bx,Riccioni:2003ga} for the D9-brane case, and also makes this result more general including other cases, like the D3-brane.
For our purpose to find the origin of the Volkov-Akulov dynamics with a single goldstino, corresponding to a single nilpotent superfield in our supergravity models \rf{ModelN}  we find it convenient to study  and to compare the cases of a  single D3-brane versus a single anti-D3-brane on top of an O3-plane. Hopefully, the phenomenon which we describe here  will be preserved in a more realistic string theory setting with many coincident branes, fluxes, curved geometry and with an account of the volume of the compactified manifold.

The outline of the paper is as follows: In section \ref{sec:D3action} we present the classical $\kappa$-symmetric D3- and anti-D3-brane actions in the flat supergravity background. In section \ref{sec:D3case} we discuss the issue of a compatibility of orientifolding with $\kappa$-symmetry gauge-fixing, following earlier studies in \cite{Bergshoeff:2005yp}. We also derive in that section the DBI and the WZ actions for D3-/anti-D3-brane with account of orientifolding and show that they both have the same fermion parts, given by the Volkov-Akulov goldstino action. Therefore, depending on the sign in the orientifolding condition, either the D3- or anti-D3-brane action vanishes whereas the other one acquires a VA goldstino action. We also discuss a possible modification of this construction in case that the flat background is replaced by a CY$_3$ compactification.
In Appendix \ref{app:A} we describe the generic case of a D$p$- or anti-D$p$-brane with the corresponding orientifold projection, and show how, in general, one finds that the WZ term upon orientifolding becomes the VA action.

\section{Classical actions for D3 and anti-D3 branes}\label{sec:D3action}

A detailed description of classical IIB D$p$-branes is given in Appendix A.1 of \cite{Bergshoeff:2013pia} and we are using the notation of this paper.
The $\kappa$-symmetric D3-brane action with $q=1$ and anti-D3-brane action with $q=-1$, in a flat background geometry consists of the Dirac-Born-Infeld-Nambu-Goto
term $S_{\rm DBI}$ and Wess-Zumino term $S_{\rm WZ}^{(q)}$ with the world-volume coordinates $\sigma^{\mu}$ $(\mu =0,\dots,3)$: \footnote{For ease of presentation we rescale the DBI and WZ term by the inverse brane tension $1/\tau_p = (2\pi)^p {\alpha'}^\frac{p+1}{2}$.}
\begin{equation}
\label{actiongeneralD3}
S_{\rm DBI} +S_{\rm WZ}^{(q)} =  - \int \rmd^{4} \sigma\, \sqrt{- \det (G_{\mu\nu} + {\alpha'} {\cal F}_{\mu\nu})} +q \int \Omega_{4} \,.
\end{equation}
Here the longitudinal and transverse coordinates are
\begin{equation}
X^m =\{X^{m^\prime}, \phi^I\}\,, \qquad  m^\prime =0,1,2,3\,,\quad  I=1,\dots,6\,,
\end{equation}
where $m^\prime$ refers to the 4 worldvolume directions and $I$ refers to the $6$ transverse directions and
\begin{eqnarray}
G_{\mu\nu} &=& \eta_{m^\prime n^\prime} \Pi_\mu^{m^\prime} \Pi_\nu^{n^\prime} +\delta_{IJ} \Pi_\mu^I\Pi_\nu^J \ ,\nonumber \\
[.2truecm]
\Pi_\mu^{m^\prime} &=& \partial_\mu X^{m^\prime} - \bar\theta \Gamma^{m^\prime} \partial_\mu \theta \ , \qquad \Pi_\mu^I = \partial_\mu\phi^I-
\bar\theta \Gamma^I \partial_\mu \theta\,.
\end{eqnarray}
The $\phi^I$ are the scalars on the D3-brane that control its position in the six transverse directions, and the Born-Infeld field strength ${\cal F}_{\mu\nu}$ is given by
\begin{equation}
{\cal F}_{\mu\nu} \equiv F_{\mu\nu} - b_{\mu\nu} \,, \qquad
b_{\mu\nu} = {\alpha'}^{-1} \bar{\theta} \sigma_3 \Gamma_{m}\partial_{\mu}\theta\left(\partial_{\nu} X^{m} -\frac{1}{2}  \bar{\theta}\Gamma^{m}\partial_{\nu}\theta\right)-\left( \mu\leftrightarrow \nu \right)\, .
\end{equation}
Finally, $\Omega_{4}$ is a particular 4-form \cite{Cederwall:1996pv,Aganagic:1996nn,Bergshoeff:1997kr}. Here we will describe it using the formalism in the flat supergravity background in \cite{Aganagic:1996nn,Kamimura:1997ju,Simon:2011rw}. Namely, we define a closed $5$-form
\be
I_{5} \equiv d \, \Omega_{4}=  d\bar \theta T_3 d\theta\,,
\label{I5}
\ee
where wedges products are implicit and the 3-form
\be
T_3=    \sigma^1{\cal F} \, \hat \Gamma + i \sigma^2  {\hat \Gamma^{3}\over 3!}\,,
\label{T3}\ee
depends on the matrix-valued 1-form\footnote{The plus sign for the second term in the 1-form is explained on page 5 in the first reference in \cite{Aganagic:1996nn}.}
\be
\hat \Gamma =\Gamma_m \Pi^m= \Gamma_m (dX^m+ \bar \theta \Gamma^m d\theta)\,.
\ee
We have also introduced the pull-backs of the flat matrices $\Gamma ^m$ to the world-volume:
\begin{equation}
\hat \Gamma_\mu \equiv \Pi_\mu{}^m \Gamma_m\,,\qquad  \hat{\Gamma }^\mu \equiv G^{\mu \nu }\hat \Gamma_\nu = \Pi^\mu _m\Gamma ^m\,,\qquad \Pi ^\mu _m= G^{\mu \nu }\Pi_\nu^n \eta_{mn}\,,
\end{equation}
where $G^{\mu \nu }$ is the inverse of $G_{\mu \nu }$. They satisfy the Clifford algebra relations
$
\hat{\Gamma }^\mu\hat{\Gamma }^\nu+\hat{\Gamma }^\nu\hat{\Gamma }^\mu=2G^{\mu \nu }
$
and
$
\Pi^\mu_m \Pi^m_\nu =\delta _\nu^\mu$.
The brane action \eqref{actiongeneralD3} has a global supersymmetry under which
$
\delta_\epsilon \Pi^m=0
$
and
$
\delta_{\epsilon}\mathcal{F} = 0
$.
Besides the global supersymmetry the action is also invariant under a local $\kappa$-symmetry (presented in details in our notation in Appendix A.1 of  \cite{Bergshoeff:2013pia} and in eq. \rf{sym} in this paper). The $\kappa$-symmetry
\begin{eqnarray}
\delta_{\kappa} \theta = (\mathbb{1}+q \Gamma)\kappa\,,
\end{eqnarray}
is defined in terms of the hermitian traceless product structure $\Gamma$ with $ \Tr\Gamma = 0\,,  \Gamma^2 = 1$. Note that in the standard $\kappa$-symmetry gauge taken in \cite{Aganagic:1996nn,Kallosh:1997aw,Bergshoeff:2013pia}
\be
(\mathbb{1}\pm \sigma^3) \theta=0\,,
\label{zeroWZ}
\ee
the WZ term \rf{I5} vanishes since \rf{T3} involves the off-diagonal $\sigma^1$ and $\sigma^2$. The gauge-fixed action of the D3- and anti-D3-brane is the same and is given in eqs. (85)-(88) in \cite{Aganagic:1996nn}.

\section{D3- and anti-D3-brane with orientifolding  }\label{sec:D3case}

The relation between orientifold truncation and gauge-fixing $\kappa$-symmetry for a D3-brane was discussed in detail in \cite{Bergshoeff:2005yp}. An orientifold action requires that
$
(\mathbb{1}-\Gamma_{\cal O}) \theta=0
$.
The gauge-fixing condition for $\kappa$-symmetry can be given in the form
$
(\mathbb{1}-\Gamma_{\kappa}) \theta=0
$.
In order for these two conditions to be compatible we need that
\be
[\Gamma_{\cal O}, \Gamma_{\kappa}]=0\,.
\ee
The O3 orientifolding studied in \cite{Bergshoeff:2005yp} for the D3-brane is defined by
$
\Gamma_{\cal O}= i \sigma^2 \Gamma_{0123}\,
$.
Thus, the general gauge-fixing condition for a D$p$-brane \rf {zeroWZ} with $\Gamma_{\kappa} = \mp \sigma^3$ which leads to a vanishing WZ term is incompatible with the O3 orientifold projection since
$
[\Gamma_{\cal O}, \Gamma_{\kappa}]\neq0
$
and the gauge-fixing for which the WZ term vanishes cannot be used.

To describe the KKLT physics we would like to demonstrate the emergence of the supersymmetric fully non-linear VA fermion action on the anti-D3-brane, and the absence of such a fermion action on the D3-brane under a certain choice of orientifolding.

We start with the action \rf{actiongeneralD3} and impose the supersymmetric truncation constraint
\be
(\mathbb{1} + q \tilde \Gamma)\theta=0\,,
\label{Orientifold}
\ee
together with
\be\label{eq:truncation}
{\cal F}_{\mu\nu}=0\,,  \qquad     \Pi_\mu^I = \partial_\mu\phi^I-\bar\theta \Gamma^I \partial_\mu \theta =0\, .
\ee
Our $\kappa$-symmetry matrix $\Gamma$ in $ \delta_{\kappa} \theta = (\mathbb{1} + q \Gamma)\kappa $ then simplifies significantly and becomes $\tilde \Gamma$ defined as follows
\be
\tilde \Gamma \equiv  \Gamma|_{(\mathbb{1} + q \tilde \Gamma)\theta={\cal F}_{\mu\nu}=\Pi_\mu^I=0} = \sigma^3 \sigma^1 {1\over 4!}
\varepsilon^{\mu_1\dots \mu_{4}}\hat \Gamma_{\mu_1\dots \mu_{4}}=  \sigma^3 \sigma^1 \Gamma^{D3}_{(0)}\,,
\label{tilde}
\ee
with $\Gamma^{D3}_{(0)}= \Gamma^{0123}$.
We have four 1-forms $\Pi^{m^\prime}=dX^{m^\prime}+ \bar \theta^1 \Gamma^{m^\prime} d\theta^1 + \bar \theta^2 \Gamma^{m^\prime} d\theta^2$, where ${m^\prime}$ are the 4 world-volume directions and  where spinors have been restricted by the condition \rf{Orientifold}. The restricted 1-forms are
\be\label{eq:Es}
E^{m^\prime}= dX^{m^\prime}+ {1\over 2}\bar \theta \Gamma^{m^\prime} (\mathbb{1}- q \tilde \Gamma) d\theta = dX^{m^\prime}+ \bar \lambda \Gamma^{m^\prime} d\lambda\,.
\ee
Here $\lambda = \sqrt 2 \, \theta^1$ is a 16-component spinor.
When all constraints are taken into account we find that the DBI action takes the form
\be
\label{actionDBI-Restr}
S_{\rm DBI}|_{{(\mathbb{1} + q \tilde \Gamma)\theta={\cal F}_{\mu\nu}=\Pi_\mu^I=0}} =  -
\int \rmd^{4} \sigma\, \sqrt{- \det G_{\mu\nu} } = - {1\over 4!}\int E^{m_0'} \wedge ...\wedge E^{m_{3}'} \varepsilon_{m_0'...m_{3}'}= - \int \det E\,.
\end{equation}
The fact that the DBI action reduces in this limit to the VA action has been known for a long time \cite{Kallosh:1997aw} and recently confirmed in \cite{Bergshoeff:2013pia}.

Now we will study the WZ term when the orientifold projection \rf{Orientifold} together with \eqref{eq:truncation} is taken into account.
This illustrates the general argument in the D$p$-brane case given in Appendix \ref{app:generalDp}. We start with
\be
q \tilde I_{5} \equiv q d \, \tilde \Omega_{4}=  d\bar \theta  q \tilde  T_3 d\theta\, \quad \text{with} \quad \tilde {T}_3 =   \sigma^3  \, \sigma^1 {(E^{m'} \Gamma_{m'})^{3}\over 3!}\,,
\ee
so that we get
\be\label{I3+2}
q \tilde I_{5} =  - E^{m_1'} \wedge E^{m_2'}\wedge E^{m_{3}'} \wedge d\bar \theta \,   q \sigma^3  \, \sigma^1 {1\over 3!} \Gamma_{m_1'm_2' m_3'} d\theta  \,.
\ee
Now we use the identity
\be
\Gamma_{m_1'm_2' m_3'}=  \varepsilon_{ m_1'm_2' m_3' m_0'}   \Gamma^{m_0'}\Gamma_{(0)}^{D3}\,,
\ee
and obtain
\be\label{I3+2new}
q \tilde I_{5}=  - {1\over 3!} \varepsilon_{ m_1'm_2'm_3' m_0'} E^{m_1'} \wedge E^{m_2'}\wedge E^{m_{3}'}\wedge d\bar \theta   \Gamma^{m_0'}  q  \sigma^3  \, \sigma^1 \Gamma_{(0)}^{D3} d\theta  \,.
\ee
Using \rf{tilde} we can rewrite this as follows
\be\label{I3+3new}
q \tilde I_{5} =  - {1\over 3!} \varepsilon_{ m_1'm_2'm_3' m_0'} E^{m_1'} \wedge E^{m_2'}\wedge E^{m_{3}'}
\wedge d\bar \theta   \Gamma^{m_0'}  q \tilde \Gamma  d\theta  \,.
\ee
Thus we get using \rf{Orientifold} and \eqref{eq:Es}
\bea\label{O3+2new++}
q \tilde I_{5} \equiv q d \, \tilde \Omega_{4} &=&  {4\over 4!}  \varepsilon_{ m_1'm_2' m_3' m_0'}   E^{m_1'} \wedge E^{m_2'}\wedge E^{m_{3}'}  \wedge dE^{m_0'}\cr
&=& - {1\over 4!}  \varepsilon_{ m_1'm_2' m_3' m_0'}  d( E^{m_1'} \wedge E^{m_2'}\wedge E^{m_{3}'}  \wedge E^{m_0'}) \,.
\eea
This can be integrated to
\be
q \tilde \Omega_{4}|_{{(\mathbb{1} + q \tilde \Gamma)\theta={\cal F}_{\mu\nu}=\Pi_\mu^I=0}}=  - {1\over 4!}   \varepsilon_{m_0' m_1'm_2'm_3'} E^{m_0'} \wedge E^{m_1'} \wedge E^{m_2'}\wedge E^{m_{3}'}  =-  \det E\,,
\ee
and we learn that our WZ term of the D3-brane or anti-D3-brane under the restrictions \rf{Orientifold} and \eqref{eq:truncation} becomes the VA action. It adds to the DBI term.

If we would use, instead, the constraint $ (\mathbb{1}-q\tilde \Gamma)\theta=0$, it would lead to a cancellation between the DBI and the WZ terms since
\be
q \tilde \Omega_{4}|_{{(\mathbb{1}- q \tilde \Gamma)\theta={\cal F}_{\mu\nu}=\Pi_\mu^I=0}}=  {1\over 4!} \varepsilon_{m_0' m_1'm_2'm_3'} E^{m_0'} \wedge E^{m_1'} \wedge E^{m_2'}\wedge E^{m_{3}'}  =  \det E\,.
\ee
In particular, for an anti-D3-brane with $q=-1$ the constraint which doubles the action is the usual O3$^-$-plane projection condition
\be\label{anti}
(\mathbb{1}+ q \tilde \Gamma)\theta=  (\mathbb{1}-\tilde \Gamma)\theta=0 \qquad \Leftrightarrow \qquad \theta^1 = \Gamma_{0123} \theta^2\,.
\ee
The conditions \eqref{eq:truncation} arise, if we place the anti-D3-brane at a fix point locus of the orientifold projection. In this case the world volume vector field $A_\mu$ and the scalars $\phi^I$ are projected out, which leads to \eqref{eq:truncation} (see appendix \ref{sec:consistency} for the vanishing of the fermionic terms). Note, that the fermions on an anti-D3-brane on top of an O3-plane are not projected out, see for example \cite{Sugimoto:1999tx}, \cite{Uranga:1999ib}. Since the analysis in \cite{Uranga:1999ib} was made in the linear approximation, the presence of fermions in absence of bosons was qualified as breaking of all supersymmetries. However, it was stressed in for example the abstract and introduction of \cite{Uranga:1999ib} that this system is free of tachyons. Meanwhile, as our non-linear analysis shows, we agree on absence of vector and scalars on a single brane, however, we find that the fermions form a goldstino multiplet with spontaneously broken supersymmetry. This fact that D-branes break supersymmetry spontaneously is often overlooked in the string theory literature although it is clearly stated for example on page 140 of \cite{Polchinski:1998rr}, where it is also mentioned that the fermions on the brane are the goldstinos.

Thus our action of the anti-D3 brane upon orientifolding is
\be
\label{actionAntiD3}
(S_{\rm DBI}+ S_{\rm WZ}^{^{(-1)}})|_{{(\mathbb{1} - \tilde \Gamma)\theta={\cal F}=\Pi^I=0}}^{\rm anti-D3} =  -2
\int \rmd^{4} \sigma\, \sqrt{- \det G_{\mu\nu} } = - 2\int E^{0} \wedge ...\wedge E^{{3}} = - 2 \int \det E\,,
\end{equation}
where
\be
E^{m^\prime}=  dX^{m^\prime}+ \bar \lambda \Gamma^{m^\prime} d\lambda\,, \qquad m^\prime= 0,1,2,3\,.
\ee
Here $\lambda$ is a 16-component Majorana-Weyl spinor (related to $\theta^1$).
This same constraint  for a D3-brane at a fixed point locus of the orientifold involution leads to a cancellation of the WZ term and the DBI term. This cancellation is a manifestation of the fact that not only the scalars and the vector on the world volume of the D3-brane are projected out but also the fermions:
\be
\label{actionD3}
(S_{\rm DBI}+ S_{\rm WZ}^{^{(+1)}})|_{{(\mathbb{1} - \tilde \Gamma)\theta={\cal F}=\Pi^I=0}}^{\rm D3} = 0\,.
\end{equation}

\subsection{Compactification}

A detailed study of the KKLT string theory model, including D-branes in a curved background with ISD fluxes, compactified on a CY$_3$ manifold will require an additional investigation. Here we will just make some plausible comments on the situation which might be expected on the basis of the results established in this paper. We also like to mention here the relevant earlier work \cite{McGuirk:2012sb}. In section 5 of this paper the authors investigate the possibility that the anti-D3-brane in a KKLT setup breaks supersymmetry spontaneously. They furthermore conjecture that the gaugino is the goldstino. However, since the authors work in the  gauge with the vanishing WZ term, they cannot distinguish between the fermionic action of an anti-D3-brane and a D3-brane and the complicated background prevents them from obtaining conclusive results. 

The comments below go beyond the scope of this work, since we studied explicitly only the case of a single anti-D3-brane on top of an O3-plane in a flat supergravity background.

Our expression for the D3- and anti-D3-brane classical action in Sec. \ref{sec:D3case} corresponds to a dimensional reduction of the D9- and anti-D9-brane classical action on a $T^6$ when all fields are assumed to be independent of the world-volume coordinates $\sigma^4,...,\sigma^9$ and after T-dualizing on all direction of the $T^6$, see for example eq. (98) in \cite{Aganagic:1996nn} where the DBI term is given. This means that the spinors on the branes remain 32-component ones in the classical actions and have 16 component upon gauge fixing $\kappa$-symmetry or upon making a supersymmetric truncation, i.e. imposing an orientifold projection. Before discussing the compactification on a CY$_3$ manifold we would like to explain here the main feature of the Volkov-Akulov theory. The action in \rf{actionAntiD3} can be shown to have a non-linear supersymmetry in a gauge where $X^{m^\prime}= \delta_\mu ^{m^\prime} \sigma^\mu$, see for example Appendix A in  \cite{Bergshoeff:2013pia}. In this gauge
\be
E^{m^\prime}|_{X^{m^\prime}= \delta_\mu ^{m^\prime} \sigma^\mu}=  d \sigma^\mu \delta^{m^\prime}_\mu+ \bar \lambda \Gamma^{m^\prime} d\lambda\,, \qquad m^\prime= 0,1,2,3\,.
\ee
The corresponding non-linear supersymmetry of the action acting on the fermion field $\lambda(\sigma)$ is given by the global parameter $\zeta$
\be
\delta_\zeta \lambda(\sigma)  = \zeta + \bar \lambda (\sigma) \Gamma^{\mu} \zeta \, \partial _\mu \lambda(\sigma)\,.
\label{susy}\ee
The first constant term shows that the supersymmetry is spontaneously broken, the second term is quadratic in fermions living on the brane.

A beautiful feature of the VA action is that one can present the symmetries of the theory in a much nicer way before gauge-fixing $X^{m^\prime}= \delta_\mu ^{m^\prime} \sigma^\mu$. The manifest supersymmetry  of the action \rf{actionAntiD3} in a form with $E^{m^\prime}=  dX^{m^\prime}+ \bar \lambda \Gamma^{m^\prime} d\lambda$ is a superspace type transformation in which the fermionic coordinate $\lambda(\sigma)$ is shifted by a global spinor $\zeta$ and the bosonic coordinates $X^{m^\prime}(\sigma)$ transform to compensate this shift
\be
\delta\lambda(\sigma) =\zeta\, ,  \qquad \delta X^{m^\prime}(\sigma)=  \bar \zeta \Gamma^{m^\prime} \lambda(\sigma)\,.
\label{superspace}\ee
Note that this superspace-type symmetry \rf{superspace} explains that the second term in \rf{susy} is just a compensating, field dependent, general coordinate transformation with a parameter
$  \xi^{\mu}_{\zeta}= \bar \lambda(\sigma)  \Gamma^{\mu} \zeta$. Note that so far we have a 16-component spinor $\lambda(\sigma)$ as well as a 16-component global supersymmetry parameter $\zeta$. This form of the VA action and its symmetries, before we gauge fix $X^{m^\prime}$, are most suitable for the discussion of the compactification on a CY$_3$ manifold.

Let us now present the sixteen component spinor $\lambda(\sigma)$ as three four dimensional spinors $\lambda^i(\sigma)$, $i=1,2,3$ that transform as {\bf 3} under the $SU(3)$ holonomy (similarly to the complex scalars $\varphi^i (\sigma)= \phi^{2i-1} + i \phi^{2i}$) and one spinor $\lambda^0(\sigma)$ that is an $SU(3)$ singlet. The global 16-component supersymmetry parameter $\zeta$  is also split into a singlet $\zeta^0$ and a triplet $\zeta^i$ under $SU(3)$. For a CY$_3$ manifold the concept of a global spinor has to be replaced by a covariantly constant spinor. Only the singlet $\zeta^0$ is covariantly constant whereas the triplets are not, see for example \cite{Jockers:2005pn}. Then the above transformations \rf{superspace}, with only the four component covariantly constant spinor $\zeta^0$ allowed, become
\be
\delta\lambda^i(\sigma) =0 \, ,  \qquad \delta\lambda^0(\sigma) =\zeta^0\, ,  \ \qquad \delta X^{m^\prime}(\sigma)=  \bar \zeta^0 \Gamma^{m^\prime} \lambda^0(\sigma) \ .
\label{superspace0}\ee
We now observe that if we do not truncate the triplet spinors on the brane $\lambda^i(\sigma)$, then the $\mathcal{N}=1$ VA supersymmetry on the brane is explicitly broken. However, if the compactification on the CY$_3$ manifold together with the orientifold projection removes the $\lambda^i(\sigma) $, then we end up with a model with $\mathcal{N}=1$ VA supersymmetry where the action of the anti-D3-brane is
\be
\label{actionAntiD3cy}
(S_{\rm DBI}+ S_{\rm WZ}^{^{(-1)}})|_{{(\mathbb{1} - \tilde \Gamma)\theta={\cal F}=\Pi^I=0}}^{\rm anti-D3\,, CY_3} = - 2\int E^{0} \wedge ...\wedge E^{{3}} = - 2 \int \det E\,.
\end{equation}
where
\be
E^{m^\prime}=  dX^{m^\prime}+ \bar \lambda^0 \Gamma^{m^\prime} d\lambda^0\,, \qquad m^\prime= 0,1,2,3 \ .
\ee
This is the VA action in d=4 corresponding to spontaneously broken $\mathcal{N}=1$ supersymmetry
\be
 \delta\lambda^0(\sigma) =\zeta^0 \,, \qquad \delta X^{m^\prime}(\sigma)=  \bar \zeta^0 \Gamma^{m^\prime} \lambda^0(\sigma) \ ,
\label{superspace0'}\ee
 which is equivalent to a chiral nilpotent superfield.

If we would take another step and assume a finite volume for the CY$_3$, we would get an action for the anti-D3-brane in Einstein frame which takes into account the volume of the extra dimensions:
\be
S^{\text{anti-D3}}=  - 2  \int d^{4}  \sigma  e^{K (\rho, \bar \rho)} \det E\,,
\ee
whereas under the same conditions we find
\be
S^{\text{D3}}= 0\,.
\ee

\section{Discussion}\label{sec:discussion}

In this note we have clarified the relation between the emergence of the nilpotent supermultiplet in $d=4$ supergravity and an anti-D3-brane on top of an O3 orientifold plane. The anti-D3-brane has a Volkov-Akulov goldstino multiplet \cite{Volkov:1973ix} on its word-volume. This construction, developing the one proposed in \cite{Ferrara:2014kva}, explains how the manifestly supersymmetric effective action based on the \K\, and superpotential in \rf{ModelN} provides the supersymmetric version of the KKLT construction. The de Sitter vacua have a spontaneously broken VA supersymmetry, which in effective supergravity  can be described by a chiral nilpotent multiplet \cite{rocek} corresponding to the emergence of the VA goldstino on the world-volume of the anti-D3 brane. In application to the KKLT model our investigation was performed so far in the simplified model of a single D3- and anti-D3-brane on top of an O3-plane in the flat supergravity background. In such a case it was possible to establish a simple connection to a supergravity effective KKLT model with an additional single nilpotent chiral multiplet corresponding to the Volkov-Akulov goldstino as given in \eqref{ModelN}. However, in a more realistic case of a full string theory one should study models with many coincident branes in a curved supergravity background, including ISD fluxes and further moduli fields like the axio-dilaton and the complex structure moduli. This we postpone to future studies.

In cosmological applications the role of the nilpotent multiplet, which has only a fermion and does not have a fundamental scalar, was shown to have various advantages over the better known supergravity models with standard chiral multiplets. In the new models there is no need to stabilize the scalar of the nilpotent multiplet since it is proportional to a bilinear of the fermions and therefore does not affect the cosmological evolution. Another advantage in using the nilpotent multiplet is that it is possible to build simple supergravity models of inflation which have an exit into de Sitter vacua \cite{Kallosh:2014via}.

The issues of cosmology raised our interest to the formal aspects of the D-brane physics and we were able to derive analytically a new result here: the Wess Zumino part of the D$p$-brane action with orientifold truncation acquires the form of the Volkov-Akulov action. This includes in particular the D3-brane case. Our derivation of this general result also explains the reason why for a D9-brane it was established computationally in \cite{Bergshoeff:1999bx,Riccioni:2003ga} that the WZ term becomes the VA action when a consistent supersymmetric orientifolding is applied.

It is instructive also to mention here again the recent progress in constructing dS vacua in \cite{Kallosh:2014oja}. In these models the effective supergravity action is manifestly supersymmetric, whereas dS vacua break supersymmetry spontaneously, on solutions, as in early dS models of this type in \cite{Saltman:2004sn}. In new models in \cite{Kallosh:2014oja} it was possible to achieve the absence of tachyons and local stability of generic dS vacua.

The current universe acceleration appears to be well described by a cosmological constant. It is therefore gratifying to find various new parts of the string theory landscape with spontaneously broken supersymmetry and an abundance of dS vacua, such as the ones in  \cite{Kallosh:2014oja}, in \cite{Ferrara:2014kva,Kallosh:2014via}, and in the advanced version of the KKLT construction presented in this paper. It would be interesting to continue exploring these kind of `supersymmetric pillars' providing uplifting and local stability of dS vacua in the landscape.

\section*{Acknowledgments}

We are grateful to E. Bergshoeff, K. Dasgupta, X. Dong, S. Ferrara, A. Hebecker, B. Heidenreich, A. Linde, J. Maldacena, T. Ortin, J. Polchinski, N. Seiberg, G. Shiu, E. Silverstein, D. Sorokin, A. Van Proeyen and A. Westphal for useful conversations, and especially to S. Kachru for  stimulating and enlightening discussions of the string theory aspects of this construction. RK and TW are supported by the SITP and by the NSF Grant PHY-1316699 and RK is also supported by the Templeton foundation grant `Quantum Gravity Frontiers'. TW is supported by a Research Fellowship (Grant number WR 166/1-1) of the German Research Foundation (DFG).

\appendix

\section{Appendix: D$p$-branes and anti-D$p$-branes with orientifolding}\label{app:A}

Here we extend the analysis of orientifolding on D$p$-superbranes in IIB string theory, which was performed for the D3-brane case in the main part of the paper. The basis for this analysis is Appendix A in \cite{Bergshoeff:2013pia}. We start with the classical action for a D$p$-brane with $q=1$ and an anti-D$p$-brane with $q=-1$:
\begin{equation}
\label{actiongeneral}
S_{\rm DBI} +S_{\rm WZ}^{(q)} =  -\int \rmd^{p+1} \sigma\, \sqrt{- \det (G_{\mu\nu} + {\alpha'} {\cal F}_{\mu\nu})} +q \int \Omega_{p+1} \,.
\end{equation}
Here $G_{\mu\nu}$ is the  manifestly supersymmetric induced world-volume metric\footnote{We use a doublet $\theta^{\alpha},~\alpha=1,2,$ of 16 component Majorana-Weyl spinors of the same chirality so that $\bar{\theta}_\alpha = \{{\theta_1}^T C,{\theta_2}^T C\}$ with $C$ the charge conjugation matrix.  $\sigma_i$ as for example in (\ref{defcalFbrane}) denotes the Pauli matrices with indices ${(\sigma _i)^\alpha}_\beta$. If it is clear from the context, we will omit the $\alpha$ indices as well as the identity matrix ${\delta^\alpha}_\beta$. We also always omit the spinorial indices.}
\begin{equation}
G_{\mu\nu} = \eta_{mn} \Pi_\mu^m \Pi_\nu^n \ , \qquad \Pi_\mu^m =
\partial_\mu X^m - \bar\theta \Gamma^m \partial_\mu \theta \,,
\end{equation}
and the Born-Infeld field strength ${\cal F}_{\mu\nu}$ is given by
\begin{equation}
{\cal F}_{\mu\nu} \equiv F_{\mu\nu} - b_{\mu\nu} \,, \qquad
b_{\mu\nu} = {\alpha'}^{-1} \bar{\theta} \sigma_3
\Gamma_{m}\partial_{\mu}\theta\left(\partial_{\nu} X^{m}
-\frac{1}{2} \bar{\theta}\Gamma^{m}\partial_{\nu}\theta\right)-\left(
\mu\leftrightarrow \nu \right)\, ,
\label{defcalFbrane}
\end{equation}
where $\Omega_{p+1}$ is a  $p+1$-form
\cite{Cederwall:1996pv,Aganagic:1996nn,Bergshoeff:1997kr}. Here we will describe it using the formalism in the flat supergravity background in   \cite{Aganagic:1996nn,Kamimura:1997ju,Simon:2011rw}. Namely, we define a closed $p+2$ form in IIB theory
\be
I_{p+2} \equiv d \, \Omega_{p+1}=  d\bar \theta T_p d\theta\,,
\ee
where in IIB models with odd $p$ the $p$-form $T_p$ is
\be
T_p= \left. e^{\cal F} \sum_{l=0} (\sigma^3 )^l \, \sigma^1{\hat \Gamma^{2l+1}\over (2l+1)!}\right|_{p-\text{form}}\,.
\label{p+2}\ee
The meaning of this expression is that $e^{\cal F}$ is expanded in powers of the 2-form ${\cal F}$ and combined with powers of the 1-form $\hat \Gamma$. $T_p$ is then picking out the $p$-forms.

Here $\hat{\Gamma}$ is the following matrix-valued 1-form
\be
\hat \Gamma =\Gamma_m \Pi^m= \Gamma_m (dX^m + \bar \theta \Gamma^m d\theta)\,,
\ee
and the pull-backs of the flat matrices $\Gamma ^m$ to the world-volume are:
\begin{equation}
\hat \Gamma_\mu \equiv \Pi_\mu{}^m \Gamma_m\,,\qquad  \hat{\Gamma }^\mu \equiv G^{\mu \nu
}\hat \Gamma_\nu = \Pi
^\mu _m\Gamma ^m\,,\qquad \Pi ^\mu _m= G^{\mu \nu }\Pi_\nu^n \eta_{mn}\,.
 \label{defhatGamma}
\end{equation}
The action \eqref{actiongeneral} has a global supersymmetry, local $\kappa$-symmetry, general coordinate symmetry and a $U(1)$ gauge symmetry:
\bea
\delta \theta &=& \epsilon + (\mathbb{1} + q \Gamma)\kappa+\xi^\mu \partial_\mu \theta\,,\cr
\delta X^M &=& -\bar{\theta} \Gamma^M \epsilon + \bar{\theta} \Gamma^M (\mathbb{1} + q\Gamma) \kappa + \xi^\mu \partial_\mu X^M\,,\cr
\alpha' \delta A_\mu &=& -\bar{\theta} \Gamma_M \sigma_3 \epsilon \partial_\mu X^M + \frac16 \bar{\theta} \sigma_3 \Gamma_M \epsilon \bar{\theta} \Gamma^M \partial_\mu \theta + \frac16 \bar{\theta} \Gamma_M \epsilon \bar{\theta} \sigma_3 \Gamma^M \partial_\mu \theta + \bar{\theta} \sigma_3 \Gamma_M (\mathbb{1}+q\Gamma)\kappa\ \partial_\mu X^M \cr
&&  -\frac12 \bar{\theta} \sigma_3 \Gamma_M (\mathbb{1}+q\Gamma) \kappa \ \bar{\theta} \Gamma^M \partial_\mu \theta -\frac12 \bar{\theta} \Gamma_M (\mathbb{1} + q\Gamma)\kappa \ \bar{\theta}\sigma_3 \Gamma^M \partial_\mu \theta + \partial_\mu \Lambda +\xi^\nu F_{\nu\mu} \,.
\label{sym}\eea
Note that this implies that
\begin{equation}
\delta_{\epsilon}\mathcal{F} = 0\,, \qquad  \delta_{\epsilon} \Pi^m=0\,.
\end{equation}
The  local $\kappa$-symmetry on fermions is given by
\begin{eqnarray}
\delta_{\kappa} \theta = (\mathbb{1}+q \Gamma)\kappa\,,
\label{kappatrans}
\end{eqnarray}
where $\kappa^{1,2}(\sigma)$, is an arbitrary doublet of Majorana-Weyl spinors of the same chirality. $\Gamma$ satisfies
$
\Tr\Gamma = 0\,,  \Gamma^2 = 1\, .
$
In the Pauli matrices basis, and acting on positive-chirality spinors $\theta^{1,2}$, $\Gamma$ is given by
\begin{equation}
\Gamma = \left(
\begin{array}{cc}
0 & \betaminus  \\
(-1)^n\betaplus & 0  \end{array} \right)\,, \label{defGamma}
\end{equation}
where $\betaplus$ and $\betaminus$ are matrices that satisfy
$
\betaminus\betaplus=\betaplus\betaminus=(-1)^n
$, with $n=(p-1)/2$.
In terms of the pull-backs, the matrices $\betaplus$ and $\betaminus$ are given by
\begin{eqnarray}
\beta_{\pm} &\equiv&\cG\,
se^{\pm\frac{{\alpha'}}{2}\mathcal{F}_{\mu\nu}\hat \Gamma^{\mu\nu}}\Gamma_{(0)}^{Dp} \equiv \cG\,\sum^{n+1}_{k=0} \frac{(\pm{\alpha'})^k}{2^k k!} \hat \Gamma^{\mu_1
\nu_1 \cdots \mu_k \nu_k }\mathcal{F}_{\mu_1 \nu_1}\cdots \mathcal{F}_{\mu_k \nu_k}\Gamma_{(0)}^{Dp}\,,
\label{defbetaplusmin}
\end{eqnarray}
and
\begin{equation}
\cG=\frac{\sqrt{\left|G\right|}}{\sqrt{\left| G + {\alpha'} \cal{F}\right|}}=
\left[ \det\left( \delta _\mu {}^\nu +{\alpha'} {\cal F}_{\mu \rho }G^{\rho \nu }\right) \right] ^{-1/2}\,.
\label{defGF}
\end{equation}
Here $se$ is  the skew-exponential function, so the expansion has effectively only a finite
number of terms.
The matrix $\Gamma_{(0)}^{Dp}$ is defined by
\begin{equation}
  \Gamma_{(0)}^{Dp} =
\frac{1}{(p+1)!\sqrt{|G|}}\varepsilon^{\mu_1\dots \mu_{p+1}}\hat \Gamma_{\mu_1\dots
\mu_{p+1}}\,,\qquad(\Gamma_{(0)}^{Dp})^2=(-1)^n\,.
\label{Gamma0Dp}
\end{equation}
For $p<9$ in expressions above we split the coordinates as follows
\begin{equation}
X^m =\{X^{m^\prime}, \phi^I\}\,, \qquad  m^\prime =0,1,\dots,p\,,\quad  I=1,\dots,9-p\,,
\label{Xmtoprimeandphi}
\end{equation}
where $m^\prime$ refers to the $p+1$
worldvolume directions and $I$ refers to the $9-p$ transverse
directions
and \begin{eqnarray}
G_{\mu\nu} &=& \eta_{m^\prime n^\prime} \Pi_\mu^{m^\prime}
\Pi_\nu^{n^\prime} +\delta_{IJ} \Pi_\mu^I\Pi_\nu^J \ ,\nonumber \\
[.2truecm] \Pi_\mu^{m^\prime} &=& \partial_\mu X^{m^\prime}
- \bar\theta \Gamma^{m^\prime} \partial_\mu \theta \ , \qquad \Pi_\mu^I = \partial_\mu\phi^I-
\bar\theta \Gamma^I \partial_\mu \theta\,.
\end{eqnarray}
Thus, $\phi^I$ are the scalars on the $p<9$ branes. When a consistent dimensional reduction of the D9-brane is performed, the $9-p$ scalars are related to $9-p$  components of the d=10 vector, namely to  $A^I$.

\subsection{$\theta^1=0$, $\theta^2=\lambda$, $X^{m^\prime}= \delta_\mu ^{m^\prime} \sigma^\mu$ gauge }

There are 32 global supersymmetries with the parameters $\epsilon^1$, $\epsilon^2$. In the gauge $\theta^1=0$, $X^{m^\prime}= \delta_\mu ^{m^\prime} \sigma^\mu$,  described in detail in \cite{Bergshoeff:2013pia} the $\kappa$ parameters and general coordinate transformation parameters $\xi^\mu(\sigma)$  become functions of $\epsilon$  and fields of the theory, so that this gauge is preserved, namely
\be
\delta \theta^1= \epsilon^1+ \kappa^1 + \beta_- \kappa^2=0
\ee
and \be
\delta X^{m^\prime}= - \bar \lambda \Gamma^{m^\prime} \epsilon^2 + (-1)^n \bar \lambda \Gamma^{m^\prime} \beta_+ \epsilon^1 + \xi^{m^\prime}=0 \ .
\ee
The gauge-fixed action has 32 global supersymmetries, 16 can be identified with deformed standard linear transformations of the vector multiplet, see eq. (A.29) in \cite{Bergshoeff:2013pia}, whereas the other 16 when acting on fermions have the form of the Volkov-Akulov non-linear transformations, see eq. (A.30) in \cite{Bergshoeff:2013pia}.

\subsection{Supersymmetric truncation ${\cal F}=0, \,\Pi^I=0\, , (\mathbb{1}\pm q  \Gamma)\theta=0$}

We define a supersymmetric truncation on the D$p$-brane and anti-D$p$-brane as follows. First we define
\be
\tilde \Gamma\equiv \Gamma|_{{\cal F}=0, \,\Pi^I=0}=(\sigma^3)^{n} \sigma^1 \tilde \Gamma_{(0)}^{Dp} \ ,
\label{tilde1}
\ee
where
\begin{equation}
 \tilde  \Gamma_{(0)}^{Dp} =
\frac{1}{(p+1)!}\varepsilon^{\mu_1\dots \mu_{p+1}} \Gamma_{\mu_1\dots
\mu_{p+1}}\,,\qquad(\tilde \Gamma_{(0)}^{Dp})^2=(-1)^n\,.
\end{equation}
There are two choices for the constraint on the spinor for actions with $\kappa$-symmetry $\delta_{\kappa} \theta = (\mathbb{1}+q \Gamma)\kappa$ which we consider. In the first case
\be
(\mathbb{1}+q \tilde \Gamma)\theta=0 \qquad \Rightarrow \qquad S= S_{DBI} + S_{WZ}^{(q)} = 2 \, S_{VA}\,,
\label{VA}
\ee
we will find that the DBI and WZ term are equal to each other and to the VA action. They therefore add up to produce the VA action.
In the second case
\be
(\mathbb{1}-q \tilde \Gamma)\theta=0  \qquad \Rightarrow \qquad S= S_{DBI} + S_{WZ}^{(q)} = 0\,,
\label{noVA}
\ee
we will find that the action is the difference between the DBI and WZ term, which each are equal to the VA action. Thus they cancel and the action vanishes. These two projections correspond to O$p$ orientifold projection and anti-O$p$ orientifold projections. The usual O$p$ orientifold projection leads to a vanishing action for the D$p$-brane and the VA action for the anti-D$p$-brane when the brane/anti-brane are located at an orientifold fixed point. The reason that the D$p$-brane has a vanishing action in this case is that all its degrees of freedom are projected by the orientifold projection. For the anti-D$p$-brane the fermionic degrees of freedom survive \cite{Sugimoto:1999tx}.

Let us consider the first case in detail. We require that the same truncation is valid for the global supersymmetry parameter as is expected for an orientifold involution
\be
(\mathbb{1} + q \tilde \Gamma) \, \epsilon=0\,,
\label{eps}
\ee
and that
\be
(\mathbb{1} + q \tilde \Gamma) \, \delta_\kappa\theta=0 \qquad  \Rightarrow \qquad (\mathbb{1} + q\tilde \Gamma) (\mathbb{1} + q\tilde \Gamma) \kappa=2 (\mathbb{1} + q\tilde \Gamma) \kappa=0\,,
\ee
and therefore
\be
(\mathbb{1} + q\tilde \Gamma) \kappa=0\,.
\ee
Thus, $\theta$, $\epsilon$ and $\kappa$ satisfy the same constraint. These conditions also serve as a gauge-fixing of the $\kappa$-symmetry.

In the second case we require that
\be
(\mathbb{1} - q \tilde \Gamma) \, \epsilon=0\
\ee
and that
\be
(\mathbb{1} - q \tilde \Gamma) \, \delta_\kappa\theta=0 \qquad  \Rightarrow \qquad (\mathbb{1} - q\tilde \Gamma) (\mathbb{1} + q\tilde \Gamma) \kappa=(\mathbb{1}- \tilde \Gamma^2) \kappa=0\,.
\ee
This condition is satisfied without constraining $\kappa$,  which means that the complete $\kappa$-symmetry gauge-fixing is not achieved. However in this case the brane action  vanishes.

\subsection{Evaluation of the action with ${\cal F}=0, \,\Pi^I=0\,, (\mathbb{1}+q  \Gamma)\theta=0$ constraint}\label{app:generalDp}

Imposing the constraint \eqref{VA}, we have $p+1$ 1-forms
\be\label{eq:Ems}
E^{m^\prime} =
d X^{m^\prime} + \bar\theta \Gamma^{m^\prime}  {1\over 2} (\mathbb{1} - q \tilde \Gamma)  d\theta = d X^{m^\prime} + \bar\lambda \Gamma^{m^\prime} d\lambda\,,
\ee
where $\lambda = \sqrt 2 \, \theta^1$.
The DBI action at  $ {\cal F}=0$, $\Pi^I=0$  and $(\mathbb{1}+ q\tilde \Gamma) \theta= 0$ becomes
\be
\label{actionDBI}
\tilde S_{\rm DBI} =  -\int \rmd^{10} \sigma\, \sqrt{- \det G_{\mu\nu} } =  -{1\over (p+1)!} \int E^{m_0'} \wedge ...\wedge E^{m_p'} \varepsilon_{m_0'...m_p'}\,.
\end{equation}
Now we look at the WZ action
\be
q \tilde I_{p+2} \equiv q d \, \tilde \Omega_{p+1}=  q d\bar \theta \tilde T_p d\theta\,,
\ee
where for $p=2n+1$
\be
\tilde {T}_p = (\sigma^3 )^n \, \sigma^1 {(E^{m'} \Gamma_{m'})^{p}\over p!}\,,
\ee
so that we get for our odd $p$
\be\label{Ip+2}
\tilde I_{p+2} \equiv d \, \tilde \Omega_{p+1}= -  E^{m_1'} \wedge ...\wedge E^{m_{p}'} \, d\bar \theta  q (\sigma^3 )^n \, \sigma^1 {1\over p!} \Gamma_{m_1'...m_p'}  d\theta\,.
\ee
We now use the following identity for odd $p$
\be
\Gamma_{m_1'...m_p'}= \varepsilon_{ m_1'...m_p'm_0'} \Gamma^{m_0'} \tilde \Gamma_{(0)}^{Dp}\,,
\label{identity}
\ee
to replace the $p$ $\Gamma$-matrices in eq. \rf{Ip+2} by their expression in \rf{identity} and obtain
\be\label{Ip+2new}
q \tilde I_{p+2} \equiv  q d \, \tilde \Omega_{p+1}=   - { (p+1)\over (p+1)!} \varepsilon_{ m_1'...m_p' m_0'} E^{m_1'} \wedge ...\wedge E^{m_{p}'}
\, d\bar \theta   \Gamma^{m_0'}  q (\sigma^3 )^n \, \sigma^1    \tilde  \Gamma_{(0)}^{Dp} d\theta\,.
\ee
Using \eqref{tilde1} and \eqref{eq:Ems} we can rewrite this as follows
\bea\label{Ip+2new+}
q \tilde I_{p+2} \equiv q d \, \tilde \Omega_{p+1}&= &- { (p+1)\over (p+1)!} \varepsilon_{ m_1'...m_p' m_0'} E^{m_1'} \wedge ...\wedge E^{m_{p}'}
\, d\bar \theta   \Gamma^{m_0'}  q \tilde \Gamma d\theta  \nonumber\\
&=&  +{ (p+1)\over (p+1)!}  \varepsilon_{ m_1'...m_p'm_0'} E^{m_1'} \wedge ...\wedge E^{m_{p}'}\wedge d E^{m_{0}'}\,.
\eea
This can be integrated to
\be\label{Op+1}
q \tilde \Omega_{p+1}|_{(\mathbb{1} + q \tilde \Gamma) \, \theta=0}=  - { 1\over (p+1)!}  \varepsilon_{m_0' m_1'...m_p'} E^{m_0'} \wedge E^{m_1'} \wedge ...\wedge E^{m_{p}'}  = -\det E\,,
\ee
and we learn that our WZ term under restrictions imposed above is proportional to the VA action!

Now we apply our findings to the D$p$-/anti-D$p$-brane action in \rf{actiongeneral}. With the supersymmetric truncation/orientifolding
\be
\label{actiongeneralTr}
\tilde S_{\rm DBI} +\tilde S_{\rm WZ}^{(q)} = \Big( - \int \rmd^{p+1} \sigma\, \sqrt{- \det (G_{\mu\nu} + {\alpha'} {\cal F}_{\mu\nu})} +q \int \Omega_{p+1}\Big)_{ {\cal F}=0,\Pi^I=0, (\mathbb{1} +q \tilde \Gamma) \, \theta=0}\,
\end{equation}
we find that the action doubles for the choice of truncation in \rf{VA}
\be
(\tilde S_{\rm DBI} +\tilde S_{\rm WZ}^{(q)})|_{ {\cal F}=0,\Pi^I=0, (\mathbb{1} +q \tilde \Gamma) \, \theta=0}
= -2 \int \det E\,.
\ee
Similarly, it vanishes for the opposite choice of truncation:
\be
(\tilde S_{\rm DBI} +\tilde S_{\rm WZ}^{(q)})|_{ {\cal F}=0,\Pi^I=0, (\mathbb{1} -q \tilde \Gamma) \, \theta=0}
= 0\,.
\ee

\subsection{Consistency of the supersymmetric truncation}\label{sec:consistency}

The action of the D$p$-brane with $p<9$ depends on scalars and vectors via the manifestly supersymmetric combinations ${\cal F}_{\mu\nu}$ and $\Pi^I$. Here we would like to show that the truncation of the scalars and the vector has to be realized via their supersymmetric combinations, as suggested in eqs. \eqref{eq:truncation} and \rf{tilde1}.

We start with the combination
\be
\Pi_\mu^I = \partial_\mu\phi^I-\bar\theta \Gamma^I \partial_\mu \theta\,.
\ee
The orientifold projection removes the scalars and we show that the fermionic term vanishes as well. We consider an (anti)-D$p$-brane extended along $01\ldots p$. The orientifolding condition given in  \rf{tilde1}  and \rf{VA} can be written as
\be\label{eq:o}
(\mathbb{1}+q \tilde \Gamma)\theta=0 \qquad \theta^1 =- q \tilde{\Gamma}_{(0)}^{Dp} \theta^2 =q	 \Gamma_{01\ldots p} \theta^2\,.
\ee
The charge conjugation matrix $C$ has the useful property
$
(\Gamma^M)^T C = -C \Gamma^M \,.
$
Now we use this and find for odd $p$ in our type IIB models (taking into account that $q^2=1$), that
\bea
\bar{\theta} \Gamma^I d\theta &=& \bar{\theta}^1 \Gamma^I d\theta^1 + \bar{\theta}^2 \Gamma^I d\theta^2\cr
&=& (\theta^2)^T \Gamma_{p}^T \Gamma_{p-1}^T \ldots \Gamma_0^T C \Gamma^I \Gamma_{01\ldots p} d\theta^2 + \bar{\theta}^2 \Gamma^I d\theta^2\cr
&=&  (\theta^2)^T C \Gamma_{p \, p-1\ldots 10} \Gamma^I \Gamma_{01\ldots p} d\theta^2 + \bar{\theta}^2 \Gamma^I d\theta^2\cr
&=& \bar{\theta}^2 \Gamma^I \Gamma_{p \, p-1\ldots 10} \Gamma_{01\ldots p} d\theta^2 + \bar{\theta}^2 \Gamma^I d\theta^2\cr
&=& - \bar{\theta}^2 \Gamma^I d\theta^2 + \bar{\theta}^2 \Gamma^I d\theta^2=0\,.
\label{Timm}\eea
Note, that the argument works also if there is an additional minus sign in the relation \eqref{eq:o}. It is instructive to explain here why
$\bar{\theta} \Gamma^{m^\prime} d\theta$, where $m^\prime= 0,...,p$ does not vanish when the same constraint on spinors is applied.
The difference lies in the fact that $\Gamma_{01\ldots p}$ commutes with $\Gamma^I$ and anti-commutes with $\Gamma^{m^\prime}$. This removes the minus in the last line of eq. \rf{Timm} so that the contributions from $\theta^1$ and $\theta^2$ instead of canceling as in the $\Gamma^I$ case, actually double. Likewise, we find that $\bar{\theta} \sigma_3 \Gamma^{m^\prime} d\theta=\bar{\theta}^1 \Gamma^{m^\prime} d\theta^1 - \bar{\theta}^2 \Gamma^{m^\prime} d\theta^2=0$. Together with \eqref{Timm} this then implies that $\beta_{\mu\nu}=0$ (cf. \eqref{defcalFbrane}). Our orientifold projection that removes the vector fields therefore leads to a vanishing of the supersymmetric version of the vector field strength ${\cal F}_{\mu\nu}=0$.

\end{document}